\documentstyle{article}

\begin{document}

\begin{center}
{\huge\bf Dimensional Structure of Space-Time}
\end{center}

\vspace{1cm}
\begin{center}
{\large\bf 
F.GHABOUSSI}\\
\end{center}

\begin{center}
\begin{minipage}{8cm}
Department of Physics, University of Konstanz\\
P.O. Box 5560, D 78434 Konstanz, Germany\\
E-mail: ghabousi@kaluza.physik.uni-konstanz.de
\end{minipage}
\end{center}

\vspace{1cm}

\begin{center}
{\large{\bf Abstract}}
\end{center}

\begin{center}
\begin{minipage}{12cm}
The dimensional structure of space-time is investigated according to physical and mathematical methods. We show that ther are various empirical and theoretical restrictions on the number of {\it independent} dimensions of space-time, consequently there is no physical and mathematical evidence for a space-time with four {\it independent} dimensions. 
\end{minipage}
\end{center}

\newpage
This report introduces a discussion about the number of {\it independent} space-time dimensions in physics, without undermining the theory of relativity which can be applied to 2 space-time dimensions \cite{un}. The question is, whether the physical space-time possesses 4 {\it independent} dimensions, or whether a manifold with 4 {\it independent} dimensions exists globally. The question arises, since firstly it is known that all 4-dimensional field theories of physics which should actually possess vector fields with 4 independent components, but in view of "constraints" possess vector fields with 4 {\it dependent} components; which implies, as it will be discussed below, that the 4 space-time dimensions should also be dependent. Secondly since there are serious mathematical doubts about the existence and uniqueness of  4-dimensional smooth manifolds which can be considered as the desirable mathematical model of the 4-dimensional physical space-time. Since according to the recent topogical (global) investigations there is no {\it existence} and uniqueness theorem for smooth 4-manifolds \cite{kosin}. Therefore there is no mathematical evidence for the {\it global} existence of a 4-dimensional space-time manifold with 4 independent dimensions, although one {\it assumes} it locally. In the same manner there is no existence and classification theorem for simply connected compact three manifolds and the Poincare conjecture could not be proved yet. Also the Banach-Tarski paradox proves that even usuall three dimensional spaces $\sim \mathbf R^3$ possess contradictional properties and are not good candidates for the physical space from a mathematical point of view, for which there is no topologically well defined candidate at all. These facts underlines our critique of the purely {\it local} standard arguments which are used to give a scientific form to the traditional {\it assumption} of 4-dimensional space-time. 

It is known that the number of dimensions of space-time was not yet an object of rigorous experiments or theoretical {\it proofs}, but as we show in the following part, it is {\it assumed} to be (3+1)-dimensional and is further adopted traditionally : 

According to the early {\it static view} formulated by Aristoteles, which {\it assumes} the "elements" and bodies to be primarily in their "natural absolute {\it rest positions}" \cite{arist}, the 3-dimensional space seemed to be neccessary to describe the arrangement of such bodies {\it at absolute rest}! Then it was {\it assumed} that we live in a (3+1)-dimensional space + time, where the additional {\it independent} time dimension becomes neccessary to describe the secondary motion of the same bodies. In other words the {\it assumption} that in a {\it static view} the primary property of bodies is to be at {\it absolute rest}, necessitates a space with 3 independent dimensions, as the arrangement arena of such bodies. Nevertheless the obvious motion of the same bodies, which is considered as a secondary, additional property of bodies in this view, necessitates the introduction of an additional independent dimension, the time, to describe the motion as the additional property of these bodies. Thus in view of the assumed absoluteness of primary rest positions of bodies in the static view, their 3-dimensional arrangement space was  also assumed to be absolute. Although this stand point was intimately related to the geocentric view, nevertheless the transition to the heliocentric view did not change the assumptions of (3+1)-dimensionality and of absoluteness of space and time. Note in this respect that as we show the 3-dimensional space with 3 independent dimensions manifests the absolute rest position of bodies and also the absolute space. The remarkable point is that despite of rejection of the concept of absolute space and time in view of the acceptance of the theory of relativity, the related concept of 3-dimensional space remains accepted.

Later the assumed absolute space and time became a relative space-time manifold with 4 relative but still {\it independent} dimensions, according to the theory of relativity; but the assumption of $4 = (3+1)$ indepndent dimensions for space and time, which originated in the static view, durates till our time. Thus all fundamental interactions in the standard model of physics are formulated by various four dimensional gauge field theories of gauge vector fields with 4 {\it independent} components \cite{grav}. In view of the fact that the four space-time dimensions are {\it assumed} to be {\it independent}, then the four space-time components of gauge vector fields should be also independent. Nevertheless despite these assumptions, on the one hand all four dimensional field theories possess "constraints" on their vector fields which reduce the number of the independent space-time components of these fields of astandard four dimensional field theory like the electrodynamics to only two components \cite{dircrm}. This reduction seems to result from the antisymmetric structure of the symplectic geometry of the phase space which separates the phase space in {\it two dimensional} subspaces. Obviously the number of the {\it independent} space-time components of dynamical variables in a four dimensional field theory, i. e. those with canonically conjugate momentum, is only two, because the number of the {\it independent} momentums in such a theory is two! \cite{dircrm}.
Then the "constraints" of any four dimensional field theory result directly from the consideration of the time as an independent dimension in the theory or from the resulting use of the time component of the field variable as an {\it independent} variable. The usual solution of this problem is that one simply eliminates the time component of field variable from the theory and use only the independent components as the "true dynamical varaibles" in the phase space of the theory neglecting the dependent components. Thus if we do not use the time components of such variables in the theory we would have no problems with constraints at all! But this would imply that we do not consider the time dimension as an independent dimension of our dynamical theory. This is exactly what results if we consider physical systems and their field theories from a dynamical point of view where ther is no need for the time as an independent dimension which is a relict from the static point of view in physics (see also below).

On the other hand it is known that the most basic field of physics which is responsible for our information, i. e. the photon field of electromagnetic radiation, possesses 2 independent space-time components, i. e. 2 {\it physical} degrees of freedom only! But this implies, that the number of independent dimensions of the base space-time manifold for Photon vector field, should be 2 only: Then in view of the isomorphism between the number of space-time dimensions and the number of components of a vector on this space-time, one number implies the other. Note that in any case the combination of the facts that the existence of constraints in a theory manifests the dependency of its field components and hence also the dependency of the dimensions of its base manifold; together with the fact that all 4-dimensional field theories possess constraints, shows that the 4 dimensions of the 4-dimensional space-time base manifold of these theories should be dependent. The main question is now: How many {\it independent} dimensions possess the physical space-time as the base manifold of physical theories?

There are further reasons to doubt the 4-dimensionality of physics, among them the fact that the modern field theory adopted the concept of four dimensional space+time from the Maxwellian physics without any {\it proof}. Although the Maxwellian physics itself has adopted this concept from Newtonian physics without {\it proof}. Thus the Newtonian physics adopted this concept from the medieval concepts of space and time and they were adopted from the Aristotelian physics which was based on the above mentioned static geocentric view with absolute space and time.  This happened, because permanent rigorous proofs of fundamental scientific concepts were not common until last century.

Moreover despite the euphemism, the so called four dimensional "standard model" of physics which is based on the assumption of four indepenent space-time dimensions, contains more basic, unsolved questions than answers: Among them the basic questions of "origin of mass", the "number of fermion families" and the " number of free parameters" for the spontaneous and chiral symmetry breaking in the theories of electroweak and strong interactions. Further the questions of the "quantization of gravity", the "constraints" and the "renormalization" of all these four dimensional quantum field theories. Thus in view of the fact that all of above mentioned difficulties of four dimensional quantum theories and the non-quantizability of gravitation theory, are related both directly and indirectly to their {\it four dimensionality}, e. g. in the sense that {\it dependent} dimensions automatically cause unsolvable problems in quantization. It seems that four dimensional field theories do not possess vector fields with four {\it independent} space-time components, because otherwise they would not have problems with quantization in view of "constraints" between these components.

Furthermore note for that the {\it most basic problem of four dimensional physics}is that all attempts to unify both fundamental theories of four dimensional physics, namely the quantum theory and the theory of relativity, were  unsuccessful: Therefore one tries to overcome this basic difficulty by some metatheory like string theory. Nevertheless, also this theory was unsuccessful several times, thus it is not even known how the string theory could be connected with the standard model of physics. 
Thus now it is certain that there is {\it no way} to quantize the {\it four dimensional} Einstein theory of gravitation \cite{witt}, since all efforts for the quantization of four dimensional theory of gravity were unsuccessful. Note that in view of the {\it four dimensionality} of this theory, the action function of the theory needs a dimensional constant, the so called Einstein constant. Nevertheless this dimensional constant prevents a canonical quantization of the theory \cite{witt}. Since the impossibility of the quantization of the four dimensional theory of gravity is a strong hint to doubt either the usefulness of 4-dimensionality of the theory of gravity, or the correctness of the standard quantization methods. 

The main point in this relation is that all quantization methods are constructed for the quantization of theories with {\it independent} canonical {\it variables} only, but not for quantization of theories with {\it constraints} and {\it dimensional constants}. Whereas the four dimensional field theories always possess according to their formulation on a four dimensional space-time manifold either {\it constraints}; and therefore {\it dependent} canonical variables; or they possess  a dimensional constant even more, as in the case of the four dimensional theory of gravity. Thus the question of the quantization of constrained theories is not solved \cite{eich} and there is no way to quantize a dimennsional constant. Then it is still not clarified, whether constraints should be applied as classical conditions to the classical variables of a theory, or whether they should be applied as quantum operators to the quantum states of the quantized theroy; since these two methods result in two different new theories. 

Thus the above mentioned certain impossibility of compatibility of both fundamental theories of {\it four dimensional} physics, which results in/from the impossibility of the quantization of the four dimensional theory of gravitation, shows that there should be something wrong with at least one of these theories. Consequently this fact implies that there should be something wrong with the four dimensional conception of physics with four {\it independent} dimensions. 

Hence one should now prove this yet unproved concept of the four dimensionality of space-time with respect to the {\it independency} of its dimensions. Thus as we show in the following part, all known arguments in favour of four dimensionality of space-time and the three dimensionality of space, e. g. according to the so called "Huygens principle" \cite{ehrund}, do not touch the question of the {\it dependency} or {\it independency} of space-time dimensions. Note that the main result of all these arguments in favour of the four dimensionality of space-time, is only that the physically reasonable number of dimensions for space and for space-time are the odd and the even ones, respectively. Further note also that even according to these results the $2 = (1+1)$-dimensional space-time is the most privileged candidate in view of the stability arguments \cite{ehrund}-(1), \cite{neu}. Nevertheless as we already mentioned one of the main errors of all known attempts to explain the 3 dimensionality of space and the 4 dimensionality of space-time is that they forget to consider the question of the {\it dependency} or {\it independency} of these dimensions. Hence despite of their arguments the explained number of dimensions in these works are always {\it dependent} dimensions. The main reason for this circumstance is the fact that all these explanations are based on infinitesimal quantities and local methods, e. g. on the local investigation of potentials and forces in various dimensions, the wave equation, etc. \cite{ehrund}. Whereas the true coordinate independent (invariant) characters of physical quantities and laws are those which are given by integral and global methods, in view of the fact that empirical, physical results are given by coordinate independent results which are hold {\it everywhere} and therefore should be described as {\it global} invariant values. Note in this respect that the value of the usual trace of a matrix, or the value of a scalar
product of two vectors are local invariants, since these values depend on the dimensions of matrix and vectors. Thus the dimension of matrices and vectors in a theory is a purely {\it assumed} number which depends on the {\it assumed} number of dimensions of the underlying manifold which is not founded in advance in the usual physical theories. Therefore another main error of these arguments for the number of dimensions of space and space-time is that they are local and dimension dependent statements \cite{ehrund}, since they found such numbers by methods which either already include these numbers in their {\it assumptions}, or they {\it manipulate} general results in a way that only the desired numbers appear to be relevant. 

To see these errors in detail note for example that despite the traditional view, it is simply not true that the $\displaystyle{\frac{1}{R}}$ potential is only given in the three dimensional case \cite{ehrund}, \cite{ehrund}-(1). Thus it is already known that also in the two dimensional case there is a vector potential with $(\displaystyle{\frac{x}{R^2}}, \displaystyle{\frac{y}{R^2}})$ components, $R^2 := x^2 + y^2$ \cite{ehrund}-(4). Then these two components can be combined into a two dimensional scalar $\displaystyle{\frac{1}{R}}$ potential, in the same way that one combines the three components of vector potential in the three dimensional case: $(\displaystyle{\frac{x}{r^2}}, \displaystyle{\frac{y}{r^2}}, \displaystyle{\frac{z}{r^2}}) \ \ , r^2 := x^2 + y^2 + z^2$ into the scalar potential $\displaystyle{\frac{1}{r}}$. Therefore the $\displaystyle{\frac{1}{R}}\sim \displaystyle{\frac{1}{r}}$ potential is by no means a privilege of the 3-dimensional case, thus any potential should be considered in the more general and fundamental context of differential geometry, as a component of some connection {\it 1-form} which may exist in various dimensions. Thus most of the above arguments \cite{ehrund} {\it presuppose} the existence of a 3-dimensional space which is  locally assumed to possess 3 independent dimensions, but it is not proven to possess 3 independent dimensions  globally (everywhere). 
Furthermore they embed for example the $\displaystyle{\frac{1}{R^2}}$-dimensional law of forces in a "3-dimensional" space in a special form, so that only the number $3$ becomes significant. But such an argument possesses no substance, since as we show the embeding presuppossess already the 3-dimensionality of the embeding space: Thus with the same right that these authors assumes $( F \sim \displaystyle{\frac{1}{R^{n - 1}}} )$ for forces in an $n$-dimensional space , one can assume $( F \sim \displaystyle{\frac{1}{R^{n}}} )$ and nothing changes in the world up to the fact that now the number $2$ becomes significant. Then the reason that one considers $( F \sim \displaystyle{\frac{1}{R^{n - 1}}} )$ for the force law in $n$-dimensional space, is simply that one assumes that the $( F \sim \displaystyle{\frac{1}{R^{2}}} )$-law is valid only in $3$-dimensional space which is not correct as we show here. Then as we discuss below, from the more rigorous differential geometric standpoint there is no invariant dependency of number of dimensions of space for the force law, if the force is considered as a component of field strength or curvature. Note that this fact is in best agreement with the common physical concept of force according to which the force is given by the gradiant of potential which can be considered as a component of a connection 1-form in view of the $\displaystyle{\frac{1}{L}}$-dimensionality of such a component (see below). Then also the curvature 2-form can be considered as the exterior differential of a connection 1-form. The same type of {\it presupposition} of 3-dimensional space is true for the potential arguments in the literature \cite{ehrund}. The reason is simply that all these arguments are Circulus vitiosus and the assumption of the 3-dimensionality of space is already involved in the arguments. 
Note also that we are not against the imbeding trick if it can solve problems, but these authors forget about the dependency or independency of the dimensions of the imbeding space and whether such an imbeding space exists as a global manifold at all. In other words the vived 3-dimensionality of objects does not prove the independency of these three dimensions from each other in view of the compactness of these objects. Thus as it is discussed later there is no rigorous possibility {\it to distinguisch three stable independent axes and hence dimensions within the compact solid objects in view of the inner dynamics of their constituting atoms and subatomic particles}. This brings us to the third related main error of these arguments, that is their origination from the traditional {\it static view} which neglects the permanent dynamics, i. e. the permanent presence of motion and forces, in the physical world. 
Thus all real physical axes and directions within or with respect to physical bodies are {\it curved} axes in view of the permanent presence of gravitational interaction which curves such material axes! Hence these directions and resulting dimensions can {\it not} be considered to be {\it independent} from each other in the same manner as the rectlinear cartesian axes. Thus not only that the construction of real cartesian axes is impossible according to the gravitation of matter, but they are even indescribable in a rigourus mathematical sense, in view of the non-integrability of {\it general three body problem} in mathematics (see below). In other words the static view and all related aspects including most of the above mentioned arguments in favour of the 3-dimensionality of space and 4-dimensionality of space-time are not only imaginary and unphysical, but they are even mathematically unprovable. 

Furthermore note that in all dimensional arguments according to the Wien's law,  Stefan-Bolzmann law and the specific heats relation for one atomic gases \cite{ehrund}, the {\it assumption} of a three dimensional space or a four dimensional space-time is the {\it presupposition} for the numbers $3$ and $4$ and $\displaystyle{\frac{2}{3}}$ in these laws, respectively. Here the parts of cause and effect are changed by authors \cite{ehrund}, since it is the assumption of 3-dimensional space or 4-dimensional space-time which results in the numbers $3$ and $4$ and $\displaystyle{\frac{2}{3}}$ in the above mentioned laws. Therefore these numbers can by no means be considered as arguments in favour of the above mentioned dimensions for space or space-time, since these are achieved by their assumptions in some previous steps. Thus these numbers $3$ and $4$ could not result from other assumptions for the space dimension or space-time dimensions, but they result if and only if one {\it presupposes} them already as the number of dimensions of space and space-time. To see this fact for example in the case of the Stefan-Bolzmann law which gives the proportionality of energy density to the fourth power of temperature $\epsilon = \sigma T^4$: Note first that it is the unfounded {\it assumption} of 4-dimensional space-time which results in the $\displaystyle{\frac{1}{L^4}}$-dimensionality of energy density in this law according to the fact that energy has the dimension of $\displaystyle{\frac{1}{L}}$ in {\it geometric units}. Thus in these units all physical dimensions are considered with respect to the dimension of length $L$ only, the time has the same dimension as length and the electric charge as well as the action are dimensionless.
Then in view of the $\displaystyle{\frac{1}{L}}$-dimensionality of temperature $(T)$, the proportionality of a 4-dimensional energy density $(\epsilon \sim \displaystyle{\frac{1}{L^4}})$ to $(T^4 \sim \displaystyle{\frac{1}{L^4}})$, is not a surprising result, but a very trivial consequence of {\it assumption} of 4-dimensionality of the underlying space-time. Moreover one can see even the origin and the growth of the number $4$ in this law from the 3-dimensional trace construction on the {\it 3-dimensional} electromagnetic pressure tensor, where the number $3$ arises directly from the trace of this 3-dimensional tensor. Then further mathematical manipulations including the integration of a related differential equation results in the discussed $T^4$-law, explicitely in view of the above mentioned {\it 3-dimensional} trace \cite{jost}. In other words, if one assumes a n-dimensional space-time, then one has a $\displaystyle{\frac{1}{L^n}}$-dimensional energy density and consequently the related Stefan-Bolzmann law would be of order $T^n$. 

In view of these facts we should be careful with the term physical law, since despite the $\displaystyle{\frac{1}{L^2}}$-law for the curvature, field strengths or forces, some of the "physical laws" are not {\it general laws} which are fulfilled in any dimension, but as we showed they are dimension dependent. The reason for the dimensional independency, i. e. the invariance of $\displaystyle{\frac{1}{L^2}}$-law for curvature in any curved space lies on the fact that the curvature is defined as a property of curved surface, i. e. always with respect to the $(L^2)$-dimensional area of such a curved surface (see below). 

Furthermore with respect to the modern arguments according to the neccessity of {\it two-component} Weyl equations and particles in favour of four dimensional space-time \cite{ehrund}-(2), note that the arguments against the space-time with more than four dimensions are correct and welcome, as like other arguments in this respect which are in favour of four or less dimensional space-time \cite{ehrund}. Nevertheless these authors forget the fact that firstly these {\it two-component} Weyl equation and 2-spinors manifests a {\it 2-dimensional} underlying space, since such a 2-spinor is the basic representation of $SO(2)$ which is the symmetry group of a 2-dimensional surface. Secondly that a general 4-dimensional space-time admits four-component Dirac equations which are {\it dependent} on each other and therefore decompose to two such two-component Weyl equations that is a hint about some 2-dimensional substructure of such a 4-dimensional space-time. Therefore even such modern arguments in favour of 4-dimensional space-time are not really in favour of the 4-dimensionality of space-time, but if they are correct then all of them point out the distinguished position of $2 = (1+1)$-dimensionality of space-time prior to its 4-dimensionality. Furthermore also these authors forget the {\it dependency} of dimensions of such a 4-dimensional space-time which admits two component Weyl equation and particles, in view of the above mentioned {\it dependency} of four components of Dirac equation. It seems that the question of the {\it dependency} of 4 dimensions is patological among the arguments which defend the 4-dimensionality of space-time. Thus not only that all arguments in favour of 3-dimensional space or 4-dimensional space-time presuppose these dimensions in some previous steps, but they always forget about the {\it question of the possible dependency} of these dimensions at all.

We investigate in the following part this question in some detail and discuss an experiment which may determine the number of independent dimensions of space empirically. We also give some mathematical-theoretical foundation for this matter. For this we investigate first the three dimensionality of space begining with the remark that the {\it vivid} "obviousness" of the "three dimensionality of space", may not confuse one, as the vividness of the "daily rotation of the sun" around the Earth does not confuse us in our heliocentric system. Then one should not confuse the vivid {\it approximate} 3-dimensionality of space, which results from the arrangment of bodies at {\it relative} rest and the {\it exact} 3-dimensionality which is related with the arrangment of bodies at {\it absolute} rest. Thus the assumed 3-dimensionality of space in physical theories is the {\it exact} 3-dimensionality which we doubt. 

Note that the origin of the modern {\it assumption} of three dimensionality of space may lie on the approximative transfer of vivid three dimensionality of macroscopic bodies to the points of space which are considered as the diminution of those bodies. Nevertheless again the question of the {\it independency} of three dimensions in macroscopic bodies is not touched by this consideration at all. Therefore we should look at this question more carefully.For this we will first announce our doubt and arguments against the {\it assumption} of the four dimensionality of space-time and then we 
will show their correctness.  

We ask the question: How many {\it independent} space-time dimensions does physics need? Or in other words, how many {\it independent} space-time dimensions can be proved by physical methods? Thus as we mentioned above, {\it dependent} dimensions result in dependent components of fields and this causes fundamental difficulties with respect to quantization of these fields and related field theories, which is a serious problem in view of the fact that quantum theory is the most basic concept to describe physics. Further we mean with physical methods, in principle, a combination of empirical and theoretical methods of physics which indicate and prove the number of {\it independent} dimensions of physical space-time. Nevertheless in the first step we accept, if the question is proved by one of theses methods, i. e. either empirically or theoretically; since indeed one can be deduced from the other. Thereby we mean with the empirical method, some realizable experiment or some known physical effect which proves either:

a. The four dimensionality of space-time with four {\it independent} dimensions, or

b. The three dimensionality of space with three {\it independent} dimensions and

c. The independency of time dimension from the three space dimensions, or

d. At least one of (b.) or (c.)

Note with respect to (c.) that any measurment of time interval measures some spatial distances or measure in the form of length, radian, or wave length and it is just by convention that one "defines" such a spatial measure to be a measure of time. Therefore any measurement and measure of time depends on some measurment and measure of space and accordingly empirically "time" is a dimension which depends on space dimensions. Thus even the notion and concept of "time", in the sense of alteration, result from both {\it spatial} rotations of the Earth around its own axis and around the sun. Therefore the time may by no means be considered as an {\it independent} dimension beside the spatial dimensions and there is no empirical way, or theoretical method to prove the independency of time as it is required in (c.). Recall that also in the more precise phase space description of dynamics, the only {\it canonical variables} are the momentum- and position variables which may depend on a time {\it parameter}; but the position variables and time parameter are never considered as {\it independent} variables. Hence even in the standard structure of physics there are enough hints about the {\it dependency} of space variables and the time parameter. 
Thus as we mentioned above, only in the static view where 3-dimensional space becomes neccessary according to the assumption that the primary property of bodies is their absolute rest; time becomes neccessary as an additional independent dimension to describe the motion of bodies as an additional property. In the realistic dynamical view where the motion is the primary property of bodies and where the existence of bodies results from their stable dynamics at all, there is no need for such a "time" as an additional independent dimension describing the additional property of motion, since the stable motion of bodies is already considered by their existence. Here the time can be considered as a parameter which may replace the space dimensions by means of the velocity or acceleration.

Furthermore note that in the very general phase space description of mechanics and dynamical systems the space and time varaibles are not independent of each other, but absolutely dependent (!) and the more rigorous symplectic description of the phase space structure does not contains the time explicitely. Thus the usual implicite introduction of the time in the phase space implies the dependent character of the time obviously. Looking on these facts without a priori views it seems obvious that there is no place for an independent time dimension in the physical dynamical systems, but the independent time dimension is a relict of the statical area of the old physics which is contra productive within the more basic and general dynamical structure of the modern physics. Nevertheless as the unsolved problems of "constraints" shows, it is the basic fault of the modern physics that it contains {\it statical} parts which are related with the independent time dimension, among them the standard "electrodynamics" which contains also the electro{\it statics}. The pure dynamical part of the standard "electrodynamics" does not possesses any constraint, since the "constraints" of any four dimensional field theory result directly from the consideration of the time as an independent dimension of the theory. It is in view of the basically dynamical character of the empirical physics and measurment that it is not possible to prove empirically the independency of the time dimension, as a purely statical concept and artifact, from the space dimensions. The {\it dynamical }character of the empirical physics and measurment is characterized by the natural and permanent presence of the physical objects and {\it forces} in any physical experience and measurment.

Then one may try to prove (b) by showing that the physical world possess either 3 independent spatial dimensions only, or equivalently $(2+1)$ independent  space-time dimensions alltogether. Nevertheless recall that this case, i. e. the odd number of space-time dimensions is rejected by the neccessity of application of Huygens principle mentioned above. In other words in view of the above discussed dependency of time dimension from the space dimensions, we are left at first sight with only 3 independent space dimensions, one of which can be replaced with and considered as an independent time dimension. Nevertheless the Huygens principle argument in favour of even dimensionality of space-time rejects this possibility. Therefore the very simple empirical fact that time measuerment depends on space measurement, thwart the long arguments in favour of four independent dimensions of space-time.   

Moreover the main reason that we mean it is not possible to prove (a.), (b.) or (c.) lies in the nature of the problem, since on the one hand one has {\it assumed} these properties till now, without any attempt to prove them. In other words known experiments and methods to prove them are laking . On the other hand the structure of physics is constructed in a way that there is a fundamental {\it invariant} and inherent two dimensional structure with two degrees of freedom, in all physical phenomena. A structure which enables one to measure and describe any physical effect by means of some force, i. e. some field strength or curvature quantity which are of {\it invariant} length dimension ($\displaystyle{\frac{1}{L^2}}$) in {\it geometric units}. Thus beyond the old known fact of $\displaystyle{\frac{1}{L^2}}$-dimensionality of the curvature of a surface, one should think in this respect on electromagnetic field strengths which are given by ($E \sim \displaystyle{\frac{Q}{K . R^2}}$), or by ($B \sim \displaystyle{\frac{Q}{C . t . R}}$) where in geometric units ($Q$) is the dimensionless electric charge, ($\displaystyle{\frac{Q}{ t}} = i$) the electric current which is of dimension ($\displaystyle{\frac{1}{t}} \sim \displaystyle{\frac{1}{L}}$ ), ($R$) is the spatial distance with dimension ($L$) and ($K$) and ($C$) are some dimensionless constants. Thus also forces have to be considered as certain components of field strength or curvature and possess the same dimension: $\displaystyle{\frac{1}{L^2}}$, a fact which also is obvious from the Lorentz force equation that is discussed below . 
The reason for the {\it invariance} of $\displaystyle{\frac{1}{L^2}}$-dimensionality of these quantities is that the most rigorous mathematical concept of force and field strength, i. e. the curvature, is defined invariantly by the surface integral of curvature according to the celebrated theorema egregium of Gauss for a two-sphere \cite{gaus}. Accordingly all our arguments here are related to a two dimensional curved compact space which is topologically equivalent to a $2$-sphere, e. g. also ellipsoid, but not to a flat two dimensional space. Therefore popular arguments against two dimensional flat space (world), e. g. by S. Hawking in \cite{haw} are not relevant in the case of curved two dimensional spaces. Note in this respect that it is in view of $\displaystyle{\frac{1}{L^2}}$-dimensionality of curvature that one is constrained to introduce the reciprocal of the $L^2$ dimensional Einstein constant in the {\it four dimensional} theory of gravity to construct the neccessary $\displaystyle{\frac{1}{L^4}}$-dimensional Lagrangian density for this theory, which however prevents the quantization of the theory, as we discussed above. Then according to the assumption of the four dimensionality of space-time for the base manifold for this theory, the action of theory must be given by a $L^4$-dimensional volume integral and hence the Lagrangian density should be of $\displaystyle{\frac{1}{L^4}}$ dimension. 

Furthermore note with respect to the above mentioned {\it two dimensional structure} of physical effects, that beyond the fact that any of our planetary orbits as a gravitational effect, can be considered to lie on a {\it two dimensional surface}; all general relativistic gravitational effects, e. g. perihelion, red shift and bending of light are effects of the order $\displaystyle{\frac{M}{R}}$ where $M$ is the mass of Sun and $R$ is some distance. In other words in view of the 
$\displaystyle{\frac{1}{L}}$-dimensionality of mass in geoemetric (field theoretical) units, all these effects are of dimension $\displaystyle{\frac{1}{L^2}}$ and hence of {\it two dimensional character} according to the $L^2$ dimensionality of the area of the mentioned surface. Thus one can consider the 
$\displaystyle{\frac{1}{L^2}}$-dimensionality of gravitational effects either as curvature effects in view of the $\displaystyle{\frac{1}{L^2}}$-dimensionality of curvature.

In the same manner all directly varified qunatum effects, e. g. Aharonov-Bohm-Casher effect (ABC), flux quantization (FQ), cyclotron motion (CM) and quantum Hall effect (QHE) can be considered as two dimensional effects: Since they are given either according to the motion of electron on the {\it boundary of a two dimensional surface} (ABC) \cite{nak}, on the boundary of the {\it flux surface} (FQ), on a circle on the two dimensional cyclotron surface (CM), or they appear only on the two dimensional electron samples (QHE). Note that all these effects are related to the action of a constant magnetic field strength $B$ on a surface $(s)$, or with the electromagnetic potential $A_l$ on the boundary $(l)$ of such a surface $(s)$ which are related together via Stockes theorem: 
$\int \int \ B ds = \oint \  A_l dl$. Thus any physical effect is measured and described by some force, field strength or curvature with dimension $\displaystyle{\frac{1}{L^2}}$ which is the reciprocal of the dimension of the area of a surface (see also below). 

Note with respect to our arguments with classical gravitational effects and quantum effects that we argue in a globaly invariant manner, since the above discussed two dimensionality of all these effects are given according to the invariant integral of the curvature or field strength \cite{gaus}. Thus these arguments do not depend on any previously assumed number of dimensions, but on the invariant length dimension of curvature and field strength, which is ($\displaystyle{\frac{1}{L^2}}$) in any space-time dimension.   

Note also that in view of the fact that any physical effect is measured according to some forces, all these effects are described according to some $\displaystyle{\frac{1}{L^2}}$-dimensional relations in the sense that $( force = ... )$, e. g. the Newton force $( F = M \cdot a )$ with the acceleration $( a )$ , or the Lorentz force $( F = Q \cdot E + Q \cdot v \cdot B )$ with the velocity $( v )$ , where both sides are of length dimension ($\displaystyle{\frac{1}{L^2}}$) in view of the dimensionlessness of velocity and electric charge in geometric units; which manifests the above mentioned fact that force and field strength are of the same dimension. In other words we have here relations of the form $( \displaystyle{\frac{1}{L^2}} = \displaystyle{\frac{1}{L^2}} )$ in geometric units. {\it If the traditional hypothetical 3-dimensionality of space, or the 4-dimensionality of space-time were really involved in physical effects, then all physical effects would be described by}: $( \displaystyle{\frac{1}{L^3}} = \displaystyle{\frac{1}{L^3}} )$, {\it or} $ ( \displaystyle{\frac{1}{L^4}} = \displaystyle{\frac{1}{L^4}} )$ {\it equations}. But on the one hand there are no such effects and as we showed already all relations of  ($\displaystyle{\frac{1}{L^3}}$), or ($\displaystyle{\frac{1}{L^4}}$) dimension, are trivial (!) in the sense that they result from previous assumptions which presuppose and fix these dimensions. On the other hand all physical effects are described by: $( \displaystyle{\frac{1}{L^2}} = \displaystyle{\frac{1}{L^2}})$ equations according to the discussed foundation on forces. In this sense, it seems that the true and stable dynamical structure of physical effects are of a 2-dimensional structure which is embedded in the traditional view of physics in an {\it assumed} 3-dimensional space and 4-dimensional space-time structure. Nevertheless such an imbeding is not helpful in the scientifical sense, because it causes unsolvable difficulties in physical theories with respect to their quantization and renormalization in view of the formulation of these theories on the imbeding space-time base manifold, since they contain superfluous and {\it dependent} variables according to the superfluous dimensions of their base manifold.

It is in view of this general and invariant ( $\displaystyle{\frac{1}{L^2}}$ ) law in the measurement and description of forces, field strength or curvature that any proper description of any physical phenomena will possess some 2-dimensional structure with respect to the square of length ($L^2$) which is the dimension of the area of a surface as a 2-dimensional geometric object. If and only if the forces, field strengths and curvature which are the basic measuring and descripting quantities of physics, possess some ( $\displaystyle{\frac{1}{L^3}}$ ) law, i. e. if they are described by some ( $\displaystyle{\frac{1}{L^3}}$ ) quantities in {\it geometric units}. Then and only then we have 3-dimensional concepts in physics with 3 independent dimensions, since $L^3$ is the dimension of volume of some 3-dimensional geometrical object with assumed independent dimensions. Therefore, since any physical effects are measured and described by some $\displaystyle{\frac{1}{L^2}}$-dimen sional force, field strength or curvature, there is no empirical possibility of proving higher dimensional concepts with more than two independent dimensions by physical methods and thereby it is not possible to prove (a.), (b.) or (c.). 

The same restriction for physical methods is also given for the basic mathematical quantities, i. e. for differential forms \cite{nak}. Thus the simple differential forms in mathematics, which are not an exterior product of other differential forms, are given up to the {\it 2-forms} only. These basic quantities are the {\it 0-form} which is a scalar function, the {\it 1-form} which has vector components and the curvature {\it 2-form} which has antisymmetric second rank tensor components. These differential forms are differential geometric descriptions of the following basic physical quantities: the action function, the momentum and connection and the field strength or curvature, respectively. Thus any higher differential form, not only in physics but even in mathematics, i. e. the Yang-Mills 4-form and Chern-Simons 3-form, should be constructed by some exterior product of the above mentioned 1- and 2-forms \cite{nak}. One can see this fact in the common structure of  the "standard model" of physics, e. g. electrodynamics, where the Lagrangian is given either by an antisymmetric product of two antisymmetric second rank tensor fields which are components of electromagnetic curvature 2-form. Or as in the case of the Einstein theory of gravity the Lagrangian is given up to the root of metric which is not a differential form, by the scalar curvature which is the trace of the tensor components of curvature 2-form. In this manner any higher dimensional mathematical structure, e. g. 3- or 4-dimensional ones, would possess some 2-dimensional substructure according to their differential form structure. 

The related two dimensional or quadratic aspect of the differential geometry is that not only that according to the Poincare duality we have a most basic quadratic structure of the differential operators in the form of the identities $d^2 = d^{\dagger}{^2} = 0$, but there is a similar distinguished quadratic structure also in view of the invaraince of the rank of the differential forms under the application of the Laplace operator which is a double derivative. 

Note however that the most fundamental such two dimensional aspects is the fact that by the virtue of the Hodge-de Rham theory, there is only a theory of invaraints of differential structure over general manifolds with a $SO(2)$ symmetry which is a typical symmetry os two dimensional manifolds. Thus this theory which is responsible for the fundamental differential topological invaraints such as the Betti numbers and related characteritic quantities, does not work if the symmetry of the manifold is higher than $SO(n), n > 2$ which are the symmetries of general manifolds of higher than two dimensions. In other words on a {\it general} higher dimensional manifold like the {\it assumed} usual four dimensional space-time manifold of physical theories, there is no mathematical theory to construct such differential topologic invariants which are neccessary to consider the topological and global aspects of the manifold in a well defined manner. It is in this manner that the celebrated Einstein theory of gravity or ART and related cosmological theories remain pure local theories without any topological character and possess only contradictious 
results if they apply to answer topological questions about the space-time.  Thus if a higher dimensional manifold admits the construction of such invaraints according to its differential geometric structure by the virtue of Hodge-de Rham theory, then it should possess only $SO (2)$ symmtry and hence it should be decomposable into some general two dimensional subspaces.
The mathematical farce to calculate such invaraints of general higher dimensional manifolds, i. e. without such two dimensional subspaces, by the methode of combinatorial topology is unmasking. Thus by definition just the possibility of the use of the combinatorial topology on a manifold presupposes the property that such a manifold is decomposable into simplexes. But if so, i. e. if a manifold is decompsable to any kind of simplexe, then such a manifold is firstly no more the announced general one and secondly it consists of submanifolds which inavoidably possess less number of dimensions than the original one! In any case the term "general higher dimensional manifold" is here definitely wrong and we have to do with a very special manifold which is built up from more simpler manifolds! In other words with or without combinatorial topology one is able to construct the above mentioned neccessary topological invaraints only for those manifolds which consists of those submanifolds for which de facto the Hodge-de Rham theory is applicable, i. e. those which are  at most general two dimensional manifolds. Note further that by such a decomposability the original manifold loses its original symmetry and it decomposes into simplexes with less symmetry, i. e. de fact with at most $SO (2)$ symmtry. 

According to these facts there is a restriction from the theoretical-mathematical side on the basic quantities in mathematics which is the mathematical equivalent of the above mentioned physical restriction of basic quantities up to the field strength and curvature. Therefore it is not possible to find a proof according to the mathematical-logical methods, for the 3-dimensionality of space or for the 4-dimensionality of space-time {\it without} any 2-dimensional {\it substructure}! Then any such method would reflect the above mentioned 2-dimensional substructure of field strength/curvature 2-forms within the 3- or 4-dimensional structure and such a substructure would manifest a repetition of the 2-dimensional substructure within the higher dimensional structure which rejects the desired independency of dimensions of higher dimensional structure.

The reason why physical phenomena possess such a two dimensional structure is a different question, but it is related to the results of the Morse-Smale theorem and the KAM theorem on the structural stability of {\it two dimensional} dynamical systems \cite{abmrs}, which guarantees the stability of physical phenomena, their repetition, repeated observation and their reproducibility by physical experiments. 

Thus we mean that according to these theorems the stability should be considered as a result of dynamics or the stable motion of the objects of physical systems. Further we mean that also the vision of space which is suggested by the stable arrangements of dynamical systems, results from the structural stability of dynamical systems such as the "Earth, Sun and Moon system" or the planetary system, etc. \cite{erdsom}. 
Accordingly we mean that the existence of 3 {\it independent} space dimensions is a characteristic of "statical systems" assumed to be at absolute rest, but it is doubtful in view of real stable dynamical systems which constitute our physical world. Thus it is obvious from our discussions that the 3-dimensional space with 3 independent dimensions manifests both the absolute rest and the absolute space which are considered in modern physics as non-physical. The problem seems to lie in the acceptance of some "approximate absolute rest" in physics, which then manifests a 3-dimensional space with 3 
{\it approximately} independent dimensions. Nevertheless then this space would possess 3 {\it approximately} independent dimensions and possess 2 exactly independent dimensions only. Note however that on the one hand the standard physical theories by which we describe our knowledge about the nature use or consider 4 exactly independent space-time dimensions. On the other hand there is no known theoretical frame work to use {\it approximately} independent space-time dimensions in a field theory and the only solution from the appearent desaster in physics is to use only field theories with absolutely independent space-time dimensions.

Furthermore note that the Earth as a well accepted reference frame belongs to such a dynamical system which is permanently in a stable motion \cite{erdsom} and every laboratory on the Earth moves permanently with the Earth in an accelaratory motion. Thus we conjecture the following simple understandable and realizable empirical method to prove the fact that in dynamical or permanently moving systems, i. e. in the real physical systems, it is not possible to observe the independency of the third dimension from the other two independent dimensions: 

Therefore let us consider a {\it moving} plane as our 2-dimensional reference frame which possesses two independent dimensions. If one draws two vertical straight axes (X) and (Y) for the two independent dimensions on this frame and throws  a body from the center of the frame in the third direction, vertical to the plane. Then this body will not fall back on the center of the frame, but on some other point with (x) and (y) components with respect to theabove mentioned axes, in view of the {\it motion} of the reference frame; showing that the {\it motion} in the third direction is not independent from the {\it motion} with respect to the other two directions. One can repeat the same experiment even with some Laser beam for more accuracy, which should be reflected on a moving reflector in high enough height from the plane, but with the same motion as the reference plane. If the height is high enough, then the reflected beam will meet a point in a remarkable distance from the center of reference frame, in view of the motion of the reference plane in the time where the beam is reflected from the reflector to the plane. Nevertheless one should also recall that Laser beams would be curved in the gravitational field of the Earth and thus they are not good candidates to construct cartesian axes which should remain absolutely vertical to each other to retain the independency of axes or dimensions. 
In other words the construction of absolute straight cartesian axes which could demonstrate 3 independent dimensions in space, is not realizable in view of the permenent existence of the gravitational field of any physical body in a physical experiment. Thus any physical cartesian frame from some solid would be curved already from the origin by means of its own weight and also in view of the gravitational field of other physical bodies. Note in this respect that the construction of a cartesian frame for a 3-dimensional space with three independent dimensions is not realizable in view of the fact that the equivalent general "three body problem" which could demonstrate the stability of a dynamical system with three independent variables or in three independent dimensions, is not stable or integrable according to the laws of mechanics 
\cite{abmrs}. Hence not only that the realization of a cartesian frame is impossible, but even the theoretical description of such a frame is not given in view of the non-integrability of mentioned dynamical systems. This fact shows the impossibility of a mathematical-logical proof of the existence of a 3-dimensional space with 3 indepndent dimensions, etc.

Recall however that if the plane were in {\it abolute rest}, then by the above experiment throwing the body in the vertical third direction, the body would fall back on the center of the frame in view of the absence of motion of the reference plane. Then this would show the independence of the third dimension from the other two dimensions on the plane. In other words only in a purely {\it hypothetical} reference frame at {\it abosulte rest}, it is possible to {\it imagine} a third dimension which is independent from the other two. This is what we mean with the purely static origin of 3-dimensionality of space which is empirically and logically unprovable and unreasonable.

One may ask the question that, if one uses the Earth surface as the two dimensional reference frame for the conjectured experiment and the two vertical geodesic circles on the Earth as the two independent dimensions on this frame, then what does the third vertical direction into the Earth mean, beginning from the intersection point of this two geodesic axes on the Earth surface frame? Is it possible to take this third direction as the third cartesian axis? But then one should recall that this third direction will cross the two circle axes at the second intersection point of geodesic axes opposite to the first one, demonstrating the fact that all these axes are not axes in the cartesian sense and are not independent from each other. Also one could assume some solid cartesian frame with 3 independent vertical directions, originated at the center of the Earth: Nevertheless firstly the independency of this solid axes is not truely realized, but only assumed; secondly these solid axes are bounded in length into the Earth and cannot be extended out of the Earth without being curved in view of their own weight. 
Forgeting about the fact that such a solid cartesian frame is not given within the Earth, since all parts, atoms and particles into the Earth are moving permanently, so that one cannot {\it even} practically distinguish one straight solid axis within the Earth and by no means 3 straight vertical solid axes in the cartesian sense within the Earth. In other words the usefulness of the thickness of any real physical two dimensional frame for the third independent dimension is doubtful from beginning, in view of the approximative nature of such materialized cartesian frames and their boundedness in the length.

Accordingly, from the scientific point of view, it is not enough to assume or to consider 3-dimensional objects, but one has to prove empirically and/or theoretically whether these 3 dimensions are {\it independent} from each other, or not; and whether such a construction is dynamically stable with respect to the permanent dynamics of the physical world? Thus one can argue in contrary that, if the 3 vivid dimensions in physical objects were really independent from each other, then by the first motion of such an object, it would decay in three directions confirming the independency of these directions within the object. Then the fact that objects with vivid 3-dimensional structure are stable and hence observable, means that the vivid 3 dimensions depends on each other holding the object together. Then any physical object consists of permanently moving atoms and those from further permanently moving subatomic particles which are examples of stable dynamical systems. Therefore only the above mentioned stability of dynamical systems \cite{abmrs} can be responsible for the stability of physical objects. Thus only the number of independent dimensions or degrees of freedom which is obeyed by the stable dynamical systems can be considered as the physical measure of the dimensions of physical objects and also of the physical space-time as the relative arrangement of such objects. In this manner, since on the one hand both the integrability of {\it general} 3-body problems and the stability of dynamical systems with more than 2 independent dimensions or degrees of freedoms are not given. On the other hand since the integrability of the restricted many body problems with 2-body substructures and the stability of higher dimensional dynamical systems with 2-dimensional subsystems are given; therefore one may doubt the independency of 3-dimensions in stable physical objects and space, but one may consider these as systems with 3 dependent and only 2 independent dimensions.
  
In other words the vivid 3-dimensionality of physical objects without an exact proof of the independency of vivid dimensions can be considered as a good argument only for the dependency of these dimensions from each others.

The reasons why till now one accepts the 3-dimensionality of space with {\it assumed} 3 independent dimensions are, firstly that one confuses the real {\it relative rest} of vivid objects with respect to the Earth, with the {\it absolute rest} position of these objects according to the traditional {\it static view} which dominates yet our view. Secondly, that just in view of the dominance of static view, even scientists {\it believe} (!) that "space" is given a priori, independent of the physical bodies and that such bodies exist {\it independent} of their motion. Although even the vivid space appears only as an arrangement of stable dynamical systems of physical bodies, which should obey the above mentioned stability theorems \cite{abmrs}. Then by the confusion between the relative rest of bodies in dynamical systems with respect to each other, and the absolute rest position for these bodies, one ignores the above discussed neccessary two dimensional character of the structural stability of dynamical systems and believes in the 3-dimensional space which results just from the assumption of absolute rest position for vivid bodies. Thus although the above mentioned unsolvable problems of 4-dimensional physical theories, which are culminated in the non-compatibility of the quantum theory and the theory of relativity, indicate a fundamrntal error in the assumption of 4 independent dimensions for space-time; however the traditional habit of static view prevents us to doubt such a problematic assumption. Nevertheless if we can look critically on this assumption and we can discard this "scientific habit", we will overcome most of the fundamental difficulties of physics.

Footnotes and references


\begin{thebibliography}{100}
\bibitem{un}
Note that usually the Lorentz transformations and other relativistic relations are calculated and represented for one space dimension and the time dimension only, e. g. the Lorentz transformations according to: 
$x^{\prime}  = f (x, t)$ and $t^{\prime} = f (x, t)$. The two other space dimensions ($y$) and ($z$) usually remain untransformed: ($y^{\prime} = y$) and ($z ^{\prime}  = z$).
\bibitem{grav}
Note also that the Einstein theory of gravitation can be formulated as a gauge field theory.
\bibitem{eich}
Note that any solution of the constraint problem, e. g. in the case of QED, results in the violation of unitarity and gauge invariance of the theory, as the most neccessary properties of modern physical theories, which produce new unsolved problems. For further discussions see Ref. \cite{dircrm}.
\bibitem{neu}
Note also that a $2$-dimensional space can be reconsidered as a $(1+1)$-dimensional space-time. Therefore we write sometimes $2$-dimensional space, but we mean also the $(1+1)$-dimensional space-time.
\bibitem{gaus}
By this fundamental theorem Gauss defines the invariant measure of the curvature of a surface according to the integral of the curvature over the surface of a curved triangle, to be equal to an angle difference which is a dimensionless and hence invariant quantity in view of the dimensionlessness of the angle (see e. g. C. F. Gauss, "collected works" and sources on the history of mathematics). Accordingly the invariant dimension of curvature is $\displaystyle{\frac{1}{L^2}}$, in view of the $L^2$ dimensionality of area of surface. Therefore the potential component of a connection 1-form possesses the invariant $\displaystyle{\frac{1}{L}}$ dimension, because its differential or derivative is the curvature 2-form with the $\displaystyle{\frac{1}{L^2}}$ dimensional curvature or field strength component. Note that this fact is related with the fact that the closed two dimensional surface integral of 2-forms (curvatures) and closed one dimensional line integral of 1-forms (potentials) in any space with any dimensions are always dimensionless invariants (winding numbers, angels).
 In this manner the true invariant and space independent dimension of potential should be given by $\displaystyle{\frac{1}{L}}$ because the line differential in the related integral has $L$ dimension and the true invariant dimension of curvature, field strength or force should be $\displaystyle{\frac{1}{L^2}}$ because the area differential in the related integral has $L^2$ dimension. This is the reason that why a curvature or field strength or force possess $\displaystyle{\frac{1}{L^2}}$ dimension even in four dimensional space-time. Any other type of attachment of dimensions to physical or mathematical quantities, which does not obey this kind of attachment according to invariant integrals of such quantities would be non-invariant and hence inadequate for physical experimental and measurments.
\end{thebibliography}
\end{document}